\documentclass[pre,aps,pacs,twocolumn]{revtex4}
\usepackage{epsfig}
\usepackage{bm}
\usepackage{color}

\usepackage[latin1]{inputenc}
\newcommand{\beq}{\begin{equation}}
\newcommand{\eeq}{\end{equation}}
\newcommand{\bea}{\begin{eqnarray}}
\newcommand{\eea}{\end{eqnarray}}

\begin{document}
\title{Numerical simulations of flow reversal in Rayleigh--B\'enard convection}

\author{Roberto Benzi$^1$ and Roberto Verzicco$^2$}
\affiliation{
$^1$ Dept. of Physics and  INFN, 
 University of Rome ``Roma Tor Vergata'', Via della Ricerca Scientifica 1, 00133 Rome, Italy, 
 $^2$ DIMeG \& CEMeC, Politecnico di Bari, Via Re David 200, 70125 Bari, Italy. }

\begin{abstract}
We investigate numerically the statistical properties of the large scale flow 
in Rayleigh--B\'enard convection. By using an external random perturbation on 
the temperature field, we were able to decrease the effective Prandtl number of the
flow while keeping the Rayleigh number relatively small; this increases the
Reynolds number thus making possible the numerical investigation
of the long--term flow statistics. We also propose a simple and 
quantitative explanation for the experimental findings on the statistical 
distribution of flow reversals and reorientations.
\end{abstract}
\maketitle

Turbulent convection in a Rayleigh--B\'enard cell is 
characterized by the complex behaviour of the unstable thermal boundary layers 
and plumes which produce a large scale (turbulent) flow in the cell. 
Experimentally, it has been observed that the large scale ``wind'' may exhibit 
abrupt flow reversals \cite{Sreeny} and reorientations \cite{Xia},\cite{Ahlers},\cite{Ciliberto},
whose statistical properties have been the subject of some theoretical 
investigations \cite{Sreeny},\cite{Detlef},\cite{Benzi}. 
Up to now, there has not been any serious 
numerical simulations aimed at investigating the statistical properties of large 
scale flow in Rayleigh--B\'enard systems.
This is due to the fact that interesting statistical properties of large scale 
flow are observed (experimentally) only at relatively large Rayleigh numbers $Ra$, 
$10^8 - 10^{11}$, and after having collected data for  rather long time periods 
(up to one year \cite{Ahlers2}). From a numerical point of view, major computer 
facilities are needed in order to properly resolve large $Ra$ conditions 
\cite{Amati} and simulating these flows for thousands of large eddy turnover 
times is, at present unfeasible. 
In this Letter, by using a suitable strategy, we show how to perform
numerical simulations aimed at investigating the statistical 
properties of large scale flow in Raylaigh B\'enard convection while maintaining
the computational requirements at a reasonable level.
The flow investigated in this Letter is that developing in
a cylindrical cell of aspect ratio (diameter $d$ over cell height $h$)
$\Gamma = d/h = 1$ heated from
below and cooled from above with an adiabatic side wall. All
the surfaces boundary conditions are no--slip. $10$ azimuthally equi--spaced
numerical ideal probes are located on a circle  halfway between the plates 
at a radius $r_p = 0.2h$; the ``probes''
provide simultaneous point-wise measurements of temperature
and of the three velocity and vorticity components.
\begin{figure}
\centering
\epsfig{width=.40\textwidth,file=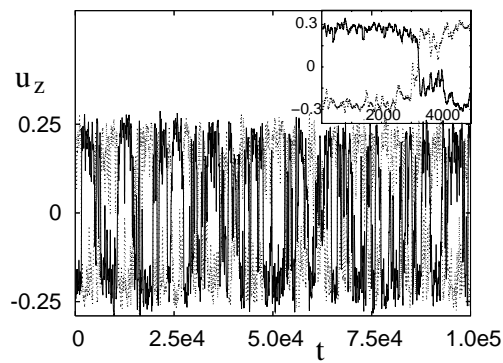}
\caption{Time evolution of the vertical velocity in one ideal probe. Many rather abrupt changes
in the signal are observed. The numerical simulation are performed using (\ref{1},\ref{2}) with an
external random forcing on the temperature equation (\ref{2}). In the inset we show the same
quantity for the same value of $Ra=6\times 10^{5}$ without the random perturbation on 
the temperature field.}
\label{fig1}
\end{figure}
The flow is solved by numerically integrating the three dimensional unsteady
Navier-Stokes equations with the Boussinesq approximation:
\beq
\label{1}
{D {\bf u} \over D t} = -\nabla p + \theta \hat{z} +
\left ( {Pr \over Ra } \right )^{1 \over 2} \nabla^2 {\bf u}, \qquad
\nabla \cdot {\bf u} = 0,
\eeq
\beq
\label{2}
{D \theta \over D t} = {1 \over (Pr Ra )^{1 \over 2}} \nabla^2 \theta,
\eeq
where $D/Dt \equiv \partial_t + {\bf u} \bullet \nabla$,
 $\hat{z}$ the unity vector pointing in the opposite direction
with respect to gravity, ${\bf u}$ the velocity vector, $p$ the pressure
(separated from its hydrostatic contribution) and $\theta$ the non dimensional 
temperature.
The equations have been made non--dimensional using the free--fall velocity 
$U=\sqrt{g\alpha \Delta h}$, the distance between hot and cold plates
$h$ and their temperature difference $\Delta = T_h - T_c$; the non-dimensional
temperature $\theta$ is defined $\theta =(T-T_c)/\Delta$ so that
$0 \leq \theta \leq 1$. The above equations have been written in a cylindrical coordinate frame
and discretized on a staggered mesh by central second--order accurate
finite--difference approximations; the resulting discretized system is solved
by a fractional--step procedure with the elliptic equation inverted using
trigonometric expansions in the azimuthal direction and the FISHPACK
package \cite{Swartz} for the other two directions. The numerical method is the same 
as that described in  
\cite{Verz96} and
\cite{Verz03} where further details of the numerical procedure
can be found. 
The numerical experiments were performed 
at $Pr=0.7$ and $Ra=6\times 10^5$ on a grid of $33\times 49\times 97$
nodes, respectively, in the azimuthal, radial and vertical directions;
this grid was found to be sufficient in \cite{Verz03} for the used flow parameters. 
\begin{figure}
\centering
\epsfig{width=.40\textwidth,file=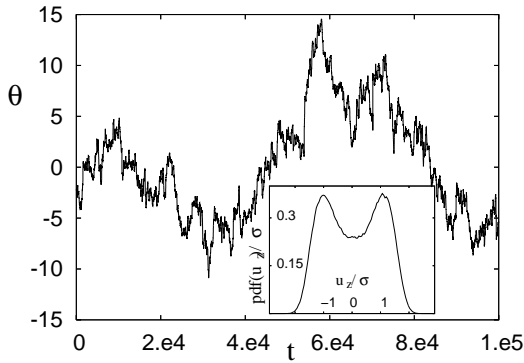}
\caption{Time evolution of the angular orientation $\Theta$ of the interface between warm rising
fluid and cold falling fluid. In the inset we show the probability distribution of the
vertical velocity  as computed in one single probe.}
\label{fig2}
\end{figure}
Owing to the relatively low value of $Ra$, large scale flow reversals or reorentations are
expected to occur on extremely long time scales, too long to achieve an acceptable statistical
convergence within a reasonable computational effort.
In order to increase the turbulent fluctuations, one can either increase the value of $Ra$
or alternatively decrease the Prandtl number (which is equivalent to increase the Reynolds $Re$
number of the system), both methods implying an increase of numerical resolution. Alternatively, we
can artificially increase thermal fluctuations by adding to the right hand side 
 of (\ref{2}) an extra forcing term $f$, $\delta-correlated$ in space and time. More precisely
the explicit expression is $f = \epsilon \sqrt{\Delta t}
\phi$ where $\epsilon$ is the amplitude of the perturbation,
$\Delta t$ is the time step size and $\phi$ is a space dependent white--noise
random number. The physical motivations for introducing $f$ are twofold: 
first of all, it has been pointed out (\cite{Benzi}) that small scale turbulent fluctuations 
can be quantitatively considered as an external random forcing acting on the large scale flow. 
Thus, one may consider $f$ as the overall effect of small ``unresolved'' turbulent motion 
for a smaller value of $Pr$. Secondly, one can
think that the effect of the external noise is to increase
thermal diffusivity and, as a consequence, to increase
$Re$ at fixed Nusselt number $Nu$  Both arguments, however, have the same physical meaning. 
In this particular example the Nusselt number resulted $Nu = 4.1 \pm 0.1$ with and without the external
noise on the temperature field, as expected, because
the Nusselt is essentially independent of $Pr$, i.e. on thermal diffusivity. 
In figure (\ref{fig1}) we
show the vertical velocity at one single ideal location as function of time for 
$\epsilon=0.1(T_h-T_c)$, while in the inset we show the same result for $\epsilon=0$. 
As we can see, a rather clear increase of the number
of reversals is observed by using the external random forcing $f$. The value
of $\epsilon \sim 0.1(T_h-T_c)$ has been chosen by trial and error, i.e. not too strong 
with respect to the deterministic dynamics and not too small to be irrelevant for the 
time scale of the reversal. As a consistency check, however, we have run similar simulations
using half and twice the above value of $\epsilon$ obtaining similar statistics for
the large scale flow.
The main simulation was run for $10^5$ time units that on account of the
cell geometry correspond to about $10^5/\pi \approx 3\times 10^4$ 
large eddy turnover times. 
Graphical inspection shows that the flow in the cell is roughly divided into
two halves, one  with a rising warm current (positive vertical velocity) and the other with
cold sinking fluid (negative vertical velocity). The two regions
 are separated by an ``interface'' where the vertical velocity is close to zero. 
Because of the cylindrical geometry, the interface has no preferred orientation and, therefore, it
is fluctuates in time. By using the same procedure described in 
\cite{Ciliberto} and \cite{Ahlers}
we can compute the instantaneous orientation  $\Theta$ of the interface. 
As a consistency check we have applied the same procedure
using either the temperature and the vertical velocity signals of the probes, always obtaining
 the same results. In figure (\ref{fig2}), we show the time behaviour of $\Theta$, while in 
the inset we show the probability distribution of the vertical velocity as obtained by 
measuring it on a single probe.
The behaviour of $\Theta$ is in close agreement with the experimental data of \cite{Ciliberto}
and \cite{Ahlers}, while the probability distribution of the vertical velocity 
shows a bimodal shape in agreement with \cite{Sreeny}. 
Next we can study the statistical properties of our large scale flow. In particular, the existence
of a bimodal distribution for the vertical velocity on a single probe, allows us to study the 
statistical properties of the (random) switching time ($\tau$) 
between the two states identified as the maxima of the probability shown in (\ref{fig2}). More precisely, we define $\tau$ 
as the time interval between two successive zero--crossings of the vertical velocity at one
ideal probe and we compute its probabity distribution $P(\tau)$.
This study was first performed by \cite{Sreeny} where an estimate of the vertical
velocity was obtained by using the correlation time of two close temperature probes. 
In \cite{Sreeny} and \cite{Ahlers}
$P(\tau)$ was found to be a power low for small enough $\tau$, while at large $\tau$ 
an exponential cutoff in $P(\tau)$ was observed. This is a rather peculiar result which 
is difficult to explain by using simple arguments based on the theory of stochastic 
differential equations or chaotic dynamics \cite{Benzi}. 
Our numerical simulations agree with the experimental results as shown in figure 
 (\ref{fig3}, circles). 
\begin{figure}
\centering   
\epsfig{width=.40\textwidth,file=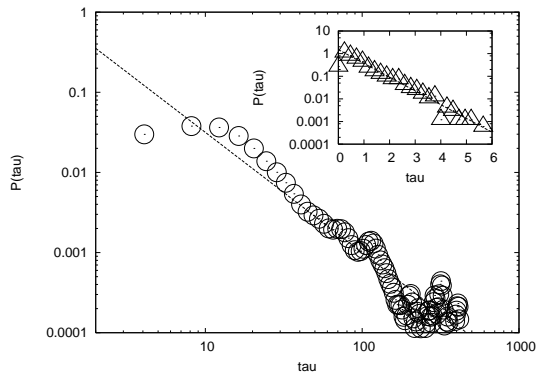}
\caption{Circles:log log plot of the probability distribution $P(\tau)$ of the random
switching time $\tau$. Note that $P$ shows a power law behaviour
shown by a straight line in the figure. Inset:
log-lin plot of the probability distribution $P_{\delta}(\tau_{\delta})$ of $\tau_{\delta}$ following \cite{Ahlers}. Note that the best fit is consistent with an exponential shape.}
\label{fig3}
\end{figure}
Following \cite{Ahlers} we also computed the probability distribution $P_{\delta}$ of 
$\tau_{\delta}$ defined as the 
time between two events where both $\delta \Theta \equiv \Theta(t+\delta t)-\Theta(t)$ and 
$d\Theta/dt \equiv \delta \Theta/\delta t$ are larger than  given
thresholds; we have used for the thresholds the same values as \cite{Ahlers} and 
have verified that the statistical results are robust with respect to the arbitrarily
selected values.  In the inset of figure \ref{fig3} we show the log-lin plot of
$P_{\delta}$ and compare it with an exponential distribution (line). 
As observed in the experimental analysis
of \cite{Ahlers}, the probability $P_{\delta}$ seems to be fitted rather accurately by 
an exponential distribution.
Figures (\ref{fig2}) and (\ref{fig3}) show our main results, i.e.  our numerical 
simulations reproduce qualitatively and quantitatively the experimental findings.
\begin{figure}
\centering
\epsfig{width=.40\textwidth,file=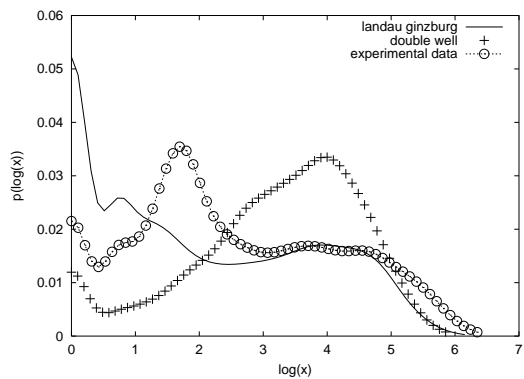}
\caption{
Probability distribution of $log(\tau)$ for the $\psi_0$ defined in (\ref{LGES}) (line) as
compared with the experimental data (circles) and the case of an overdamped brownian particle 
in a double well potential (see text for a description).
}
\label{fig4}
\end{figure}
\begin{figure}
\centering
\epsfig{width=.40\textwidth,file=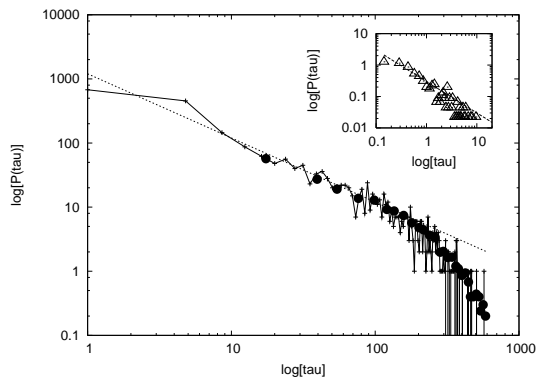}
\caption{Log-log plot of the probability distribution $P(\tau)$ of the random
switching times $\tau$ between the maxima of $w$. Small black 
circles are obtained using (\ref{model}) with
$\Theta$ a random noise correlated in time. 
Line with symbol refers to the experimental data by \cite{Sreeny}.
In the insert, we show the same quantity computed with
$\Theta$ obtained by our numerical simulations, see figure (\ref{fig2}).}
\label{fig5}
\end{figure}
Next, we want to discuss briefly some theoretical ideas which may 
help in explaining the power law behaviour of the probability distribution $P(\tau)$ 
as computed by \cite{Sreeny} and reproduced in our simulations. From a theoretical point of view,
we can state the problem as follows: Let $Q$ be a random variable whose probability distribution
is bimodal with maxima in $\pm Q_0$ and let $\tau_Q$ be the (random) switching time between 
the maxima.
Is there any ``simple'' model for which $P(\tau_Q)$ shows a power law behaviour for small $\tau_Q$?
We will show that such a model does exist and it may be relevant for thermal convection. To this aim,
let us consider the Landau-Ginzburg equation (LGE) in two space
dimensions for the field $\psi(x,y,t)$:
\beq
\partial_t \psi = m \psi -g \psi^3 + \nu \Delta \psi + \sigma \eta(x,y,t)
\label{LGE}
\eeq
where $m,g,\nu,\sigma$ are real positive variables and $\eta$ is a white noise {\it delta} correlated in
space and time. We assume $\psi$ defined on a periodic box of finite size $L$ and we indicate with $\langle .. \rangle_s \equiv
L^{-2} \int .. dxdy$ the space average. We can decompose $\psi$ in its space average 
$\psi_0 \equiv \langle \psi \rangle_s$ and a fluctuation $\phi$, i.e. $\psi \equiv \psi_0 + \phi$. Then the equation for $\psi_0$
reads:
\beq
\label{LGES}
\partial_t \psi_0 = (m-3g\langle \phi^2 \rangle_s) \psi_0 - g \psi_0^3 + \sigma_s \eta(t)
\eeq
where $\eta$ is a white noise {\it in time}. More precisely, let us consider (\ref{LGE}) defined on a regular lattice
of $N \times N$ points of spacing $\Delta x = \Delta y = L/N$. Then $\sigma_s \equiv \sigma/N^2$ and we 
assume $\sigma_s = const.$ for any $N$ (see \cite{jona} for a correct 
definition of the problem). It is worth reminding that the equilibrium probability 
distribution of (\ref{LGE}) has been widely used as a  model of second order 
phase transition in the infinite volume limit. For a suitable choice of the 
parameters $m$,$g$ and $\nu$, the term $\langle \phi^2 \rangle_s$ can be large when 
$\psi_0$ becomes small, i.e. the system at $\psi_0 \sim 0$ shows strong fluctuations as 
it is usual in second order phase transitions.  In this case, the linear term in (\ref{LGES}), 
i.e.  $(m-3g\langle \phi^2 \rangle_s)$, can occasionally change sign and the switching 
between the two maxima of the probability distribution of $\psi_0$ can be more frequent.
This is equivalent to say that the switching time is no longer controlled by the mechanism 
leading to an exponential distribution and one can observe a scale (in time) invariant 
probability distribution, i.e. a power law. 
In figure (\ref{fig4}) we show the probability distribution of  the
switching time of $\psi_0$ obtained by numerically integrating equation (\ref{LGE}), 
for $N=32$, $\nu=0.1$, $m=1$, $g=9$ and $\sigma_s = 0.05$. In particular we plot
the probability distribution of $log(\tau)$, where $\tau$ is the random switching time 
computed by using the parameter $\psi_0$.
Note that if $P(\tau)\sim \tau^{-1}$ then $P(log(\tau)) \sim const.$ In the
same figure we show the probability distribution of $log(\tau)$ for the experimental 
data by \cite{Sreeny}) and for 
the well known case of an over-damped Brownian particle in a double well potential 
with the same parameter $m$ and $g$ and
stochastically perturbed by a white noise with variance $\sigma_s$.
In the last case, we know
that the probability distribution of $\tau$ is exponential and, indeed, in figure 
(\ref{fig4}), $P(log(\tau))$ for 
the double well potential is not constant. On the other hand, $P(log(\tau))$ is 
constant for almost one decade for the numerical
simulation of (\ref{LGE}) using $\psi_0$ as order parameter.
Thus, we may tentatively state that the equation (\ref{LGE}), and in particular 
(\ref{LGES}), can be
considered as a conceptual model for the experimental results of \cite{Sreeny}). 
From a physical point of view, we remark that there exists a non trivial 
feedback between the variance of the
fluctuations {\bf in space} and the value of the order parameter $\psi_0$. 
It is now tempting to understand whether the same ``physical'' mechanism can be observed in 
our numerical simulation of the Rayleigh--B\'enard turbulence. According to our previous 
discussion, the overall picture which emerges is quite simple and agrees with the original 
description of \cite{Ciliberto}. As a first approximation we can think that the large scale flow
rotates randomly in time with some orientation $\Theta(t)$ performing a random walk in the interval
$[0,2\pi]$. The vertical velocity in a single point is almost insensitive to $\Theta$ unless the
interface is close to the ideal probe. A suitable model for $\Theta(t)$ can be build as follows.
Let us consider two variables $x_1 = r \cos(\theta)$ and $x_2 = r \sin(\theta)$ which 
satisfy the following equation
\beq
\label{xi}
\frac{dx_i}{dt} = - \frac{\partial V}{\partial x_i} + noise
\eeq
where $V \equiv - 1/2 r^2(R^2-r^2)$ and $r^2 = x_1^2+x_2^2$, i.e. $r\sim R$ for most of the time. 
We can imagine that $\theta$ is precisely the variable $\Theta$ 
describing the interface direction of the large scale wind.
Then
denoting by $w$ the value of the vertical velocity, 
a qualitative model of $w$ as a function of $\theta(t)$ is simply given by:
\beq
w = \frac{x_2}{\sqrt{x_2^2+A}} = \frac{sin(\theta)}{\sqrt{sin^2(\theta)+A/r^2}}
\label{model}
\eeq
where the parameter $A$ is a measure of the interface thickness, whose fluctuations are controlled
by $r$. Using  (\ref{model})
and our simple model, one can easily show that:
\beq
\frac{dw}{dt}= aw -(a+bA)w^3 + noise
\label{x2}
\eeq
where $a \equiv R^2-2x_1^2$ and $b = 2.$. 
Equation (\ref{x2}) has a form of an over-damped Brownian particle in a double well potential. 
However the term $a$ can change sign (exactly as in (\ref{LGES})) because of the
(strong) fluctuations of  $x_1$. In order to compute the probability distribution of 
the switching time $\tau$ between the two maxima in the probability of $w$,
we have numerically integrated equation (\ref{xi}), with $R=1$, $A=0.1$ and the 
variance of the noise equal to $0.1$.
The final result is shown
in figure (\ref{fig5}) (small black circles) and compared with the experimental 
finding of \cite{Sreeny} (line with symbol).
As we can see $P(\tau)$ is close to a power law with slope $-1$ in agreement with the experimental
findings of \cite{Sreeny}. In the same figure, we show the probability distribution 
$P(\tau)$ computed by using the approximate value $ w \sim sin(\Theta)/\sqrt{sin^2(\Theta)+A}$
 and the value of $\Theta$ obtained in the numerical simulation (see figure
(\ref{fig2})). A good agreement is observed with a power law $P(\tau) \sim \tau^{-1}$. 
Thus (\ref{model})
seems to capture the physical reason for the observed numerical and experimental results in the 
statistical properties of flow reversal. 
In summary, we have shown that the statistical properties of large scale flow in Rayleigh
B\'enard convection can be efficiently simulated by using an external random forcing in the 
temperature
field. The effect of the forcing is to increase thermal diffusivity, i.e. to decrease the Prandtl
number. Consequently, the effective Reynolds number achieved in this way is 
larger with respect to the one without external forcing. Our numerical simulations agrees qualitatively
and quantitatively with the experimental findings of \cite{Sreeny}, \cite{Ahlers} and \cite{Ciliberto}. 
Finally, we have discussed some theoretical ideas which may be able to explain
the statistical properties of flow reversals. 

We acknowledge K.R. Sreenivasan, G. Alhers, K.Q. Xia, T. Sato, L.  Kadanoff, J. Niemela, for useful discussions.

\end{document}